\documentstyle[12pt]{article}

\def\bib{ \begin{thebibliography}{99}}
\def\finbib{\end{thebibliography}}
\font\hdr=cmbx12 scaled1500
\font\chap=cmbx12 scaled1000

\def\ce{\centerline}
\newcommand{\head}[1]{\centerline{\hdr{#1}}}
\newcommand{\chapt}[1]{{~}\\ \centerline{\chap{#1}}{~}\\ }
\newcommand{\chapl}[1]{\centerline{\chap{#1}} }

\begin{document}
\def\doub{}
\def\nsz{}
\def\be{\begin{equation}\nsz}
\def\ee{\end{equation}\doub}
\def\ba{\begin{eqnarray}\nsz}
\def\ea{\end{eqnarray}\doub}
\def\next{\end{eqnarray}\begin{eqnarray}}
\def\g{\gamma}
\def\<{\left<}
\def\>{\right>}
\def\({\left(}
\def\){\right)}
\def\[{\left[}
\def\]{\right]}
\def\l|{\left|}
\def\r|{\right|}
\def\intz{\int\limits_0^\infty}
\def\.{\right.}
\def\ts{\widetilde S}
\def\tg{\widetilde G}
\def\G{\widetilde g}
\def\QED{QED${}_3~$}
\def\QEDR{QED${}_4~$}
\def\eis{e^{-is\varphi_1}}
\def\gm{\g_\mu}
\def\gn{\g_\nu}
\def\gz{\g_0}
\def\exi{e^{i\sigma_3\xi}}
\def\eet{e^{i\sigma_3\eta}}
\def\gpar{g^\parallel_{\mu\nu}}
\def\gper{g^\perp_{\mu\nu}}
\def\gmn{g_{\mu\nu}}
\baselineskip=12pt
\head{Dynamical mass generation in 2+1-dimensional }
\head{electrodynamics in an external magnetic field.}
\ce{Shpagin A.V.}
\ce{\it Kiev State University}
\ce{\it Glushkov Av.,6,252022}
{~}\\
\ce{\bf Abstract.}
The influence of a magnetic field on the mass generation
in 2+1 dimensional QED is considered.It is shown that the magnetic
field is a catalyst of the generation of a fermion dynamical mass.
The mass arises in the system with arbitrary number of fermions,
not only with $N\leq 4$, as it is in the system without
the magnetic field.
The polarization tensor
is calculated for a constant magnetic field.
\pagebreak
\chapt{Introduction.}
\doub

 It was shown in [1-4] that a constant magnetic field
is a strong catalyst of dynamical mass generation and
flavor symmetry breaking because of the interaction between
fermions. This effect is due to the dimensional reduction
($D\to D-2$) in dynamics of pairing of fermions in a magnetic field.
It is connected with restricton of the motion of charged particles
in directions that are perpendicular to the magnetic field.
The phenomenon of the catalysis was illustrated by the
Nambu--Jona--Lasinio model (NJL) in 2+1 and 3+1 dimentions,
quantum electrodynamics (QED) in 3+1 dimensions. The crucial role
of the lowest Landau level (LLL) in catalyzing the spontaneous
symmetry breaking was emphasized. The role of the LLL in the
effect of dynamical symmetry breaking is similar to the role of
the Fermi surface in the BCS theory.

Here we investigate effect of the dynamical mass generation
in 2+1 electrodynamics (\QED) in an external magnetic field.
There are N flavors in the system, initial lagrangian is massless.
This model is interesting from two points of view. At first,
\QED can serve as an effective theory for the discription
of longvawe excitations in planar systems in the condensed
matter theories \cite{physMean}.
 \QED also has properties reminiscent of QCD and
other four- dimensional gauge theories \cite{applq,QED3}.
The investigation of the Schwinger- Dyson equation for
the fermion self energy in $1/N$ expansion without magnetic
field indicates that there exists critical number of fermions
($N_c\simeq 3\div 4$) \cite{applq}, 
the chiral symetry is broken for $N<N_c$. In such a way \QED has two phases
(massive and massless) depending on parameters of the theory, as
it was in NJL and \QEDR (in ladder approximation).
But in contrast to \QEDR polarization effects must be
taken into acccount to obtain the chiral phase transition.
With turning on a strong magnetic field in \QED the dimensional 
reduction occurs and dynamical mass of fermions is generated
for arbitrary number of flavors.

This article contains three sections and appendix: in the first section
the dimensional reduction of the space is considered with
help of the Schwinger- Dyson equation, in the second section
the effects that are connected with vacuum polarization are under
consideration, in the third section the expression for a mass
of fermions is obtained. The polarization tensor for the \QED
in the constant external magnetic field is obtained in
appendix.

\chapt{1.The Schwinger-Dyson equation for the system with a
magnetic field.}

In the presence of a magnetic field the infrared region 
is responsible for the mass
generation. This justifies using
the ladder approximation because the full vertex 
$\Gamma^\mu$ is replaced by its infrared asymptotics $\g^\mu$
(Dirac matrices). Let us consider the Schwinger--Dyson (SD) 
equation in the improved ladder approximation 
(the vacuum polarization is taken into account).
\def\GZ{\gamma}
\be
G(x,y)=S(x,y)-ie^2 \int d^3z d^3t S(x,z)\g^\mu G(z,t)\GZ^\nu
 G(t,y) D_{\mu\nu}(t-z),
\label{scoor}
\ee

where $G(x,y)=-i\<0\l|T\psi(x)\bar\psi(y)\r|0\>$ is the full 
electron propagator, 
$D_{\mu\nu}(x-y)=i\<0\l|TA_\mu(x)A_\nu(y)\r|0\>$ is the full
photon propagator, $S$ is the free fermion propagator.
\be
S(x,y)=exp\({ie\over 2}(x-y)^\mu A_\mu^{ext}(x+y)\)\ts(x-y)~~,~~A^{ext}=(0,-{B\over 2}x_2,{B\over 2}x_1).
\label{free}
\ee
  The expression for $\ts$ will be written down further. The full
electron propagator can be represented as follows:
\be
G(x,y)=exp\({ie\over 2}(x-y)^\mu A_\mu^{ext}(x+y)\)\tg(x-y).
\label{full}
\ee
In terms of $r=x-y, R={x+y\over 2}$ the SD equation takes the form:
\ba
\tg(r)=\ts(r)-ie^2 \int d^3R_1 d^3r_1\ts\({r-r_1\over 2}-R_1\)
\g^\mu\tg(r_1)\GZ^\nu\tg\({r-r_1 \over 2}+R_1\)\nonumber\\
\cdot D_{\mu\nu}(-r_1) exp\(ie(r+r_1)A^{ext}(R_1)\).
\label{e1}
\ea
Transforming this into the momentum space, we obtain:
\ba
\tg(p)=\ts(p)-ie^2\int {d^2k_\perp d^2R_\perp d^2q_\perp dk_0 \over (2\pi)^5}
e^{-iq_\perp R}\ts\(p_0,p_\perp+eA(R)+{q_\perp\over 2}\)
\nonumber\\
\cdot\g^\mu\tg(k)\GZ^\nu
\tg\(p_0,p_\perp+eA(R)-{q_\perp\over 2}\) D_{\mu\nu}(k_0-p_0,k_\perp
-p_\perp-2eA(R)),
\label{e2}
\ea
where $q_\perp=(q_1,q_2)$.
Schwinger found the exact expression for the free electron propagator 
in an external magnetic field \cite{shwing}:
\ba
S&=&exp\left(ie(x-y)^\mu A_\mu^{ext}\left({{x+y}\over 2}\right)\right)
\widetilde S(x-y),\nonumber\\
\widetilde S(k)&=&-i\int ds exp\left(-ism^2+isk_0^2-isk^2{tg(eBs)\over eBs}
\right)\nonumber\\
&\cdot&(k^\mu\gamma_\mu+(k^2 \gamma^1-k^1 \gamma^2)tg(eBs))
(1+\gamma^1\gamma^2 tg(eBs)).
\label{stilde}
\ea
 According to \cite{decomp}  $\widetilde S$ can be decomposed
over the Landau level poles:
\ba
\widetilde S(k)=-exp\left(-{{k_\perp}\over{|eB|}}\right)
\sum\limits_{n=0}^\infty{{(-1)^n D_n(eB,k)}\over{m^2-k_0^2-2|eB|n}},
\label{decompose}
\ea
where
\ba
D_n(eB,k)=(m+k^0\gamma^0)\((1-i\gamma^1\gamma^2 sign(eB))
L_n\left({{2k^2_\perp}\over{|eB|}}\right)\right.\nonumber\\
\left.-(1+i\gamma^1\gamma^2 sign(eB))
L^1_{n-1}\left({{2k^2_\perp}\over{|eB|}}\right)\)
+4(k^1\gamma^1+k^2\gamma^2)L^1_{n-1}\left({{2k^2_\perp}\over{|eB|}}\right).
\nonumber
\ea
Therefore, in strong enough field,  
the sum can be reduced to only one term
($k,m_{dyn}\ll \sqrt{e|B|}$).
\be
\ts(p)=e^{-\ell^2p_\perp^2}{1\over \g^0 p_0-m}(1-i\g^1\g^2\cdot sign(B)),
\label{SbigB}
\ee
$\ell=1/\sqrt{e|B|}$ is a magnetic length (let us take $B>0$).
The main contribution comes from the LLL. And this leads to the 
dimensional reduction. It is natural to search the full
electron propagator in the form
\ba
\tg(p)=e^{-\ell^2p_\perp^2}g(p_0).
\label{fullTG}
\ea
If we substitute this into (\ref{e2}), we obtain the equation for $g$:
\ba
g(p_0)\g^0p_0=1-i\g^1\g^2-{ie^2\over 4(2\pi)^3}(1-i\g^1\g^2)
\int dk_0 \g^\mu g(k_0)\GZ^\nu g(p_0)D^0_{\mu\nu}(k_0-p_0),
\label{maineq}
\next
D^0_{\mu\nu}(p_0)=\int d^2t_\perp
e^{-\ell^2t^2_\perp/2}D_{\mu\nu}(p_0,t_\perp).
\nonumber
\ea

  Let us consider the matrix structure of the full electron propagator.
We can write the SD equation in two equvalent forms:
\ba
G(x,y)=S(x,y)-ie^2 \int d^3z d^3t S(x,z)\g^\mu G(z,t)\GZ^\nu
 G(t,y) D_{\mu\nu}(t-z),
\nonumber
\ea
or
\ba
G(x,y)=S(x,y)-ie^2 \int d^3z d^3t G(x,z)\g^\mu G(z,t)\GZ^\nu
 S(t,y) D_{\mu\nu}(t-z).
\nonumber
\ea
  Therefore the full propagator takes the form (see( \ref{SbigB}))
\be
G={1-i\g^1\g^2\over 2}\xi{1-i\g^1\g^2\over 2}
=P\cdot\xi~~,~~P^2=P,
\label{MatrG}
\ee
where $\xi$ is a matrix  $4\times 4$. In the standard representation
\ba
P\(\matrix{a & ? & b & ?\cr
	     ? & ? & ? & ?\cr
	     c & ? & d & ?\cr
	     ? & ? & ? & ?}\)=
\(\matrix{a & 0 & b & 0\cr
	  0 & 0 & 0 & 0\cr
	  c & 0 & d & 0\cr
	  0 & 0 & 0 & 0}\)\sim
\(\matrix{a & b\cr
	  c & d\cr}\).
\ea
In such a way a strong magnetic field restricts matrix structure 
of the full electron propagator to  $G\in GL(2,C)$.
Hence the equation for the full electron propagator 
reduces to the form:
\def\la{\Lambda}
\ba
\la(2\G\g^0p_0-2)\la=-{ie^2\over 4(2\pi)^3}2\la\int dk_0\g^\mu\la2\G(k_0)\la
\GZ^\nu\la2\G(p_0)\la D^0_{\mu\nu}(k_0-p_0),
\nonumber
\next
\Lambda={1-i\g^1\g^2\over 2}~,~~
g=\la 2\G\la,
\label{LGL}
\ea
where $\G$ is in general the linear combination of four matrices 
$I,\g^0,\g^3,\g^5.$
Because of identies $\la\g^2\la=\la\g^1\la=0~,~[\la,\G]=0,$ we get
\ba
{1\over \G(p_0)}=p_0\g^0+{ie^2\over (2\pi)^3}\int dk_0\g^0\G(k_0)\g^0 D^0_{00}(k_0-p_0).
\ea
This equation has the one-dimensional form what
justifies the term 'dimensional reduction'. The system of
equations for scalar functions can be obtained if we put
\ba
G=e^{-p^2_\perp\ell^2}{1\over A\g^0p_0-B}(1-i\g^1\g^2)~~,~~
\G={1\over A\g^0p_0-B},
\label{SupG}
\next
p_0\(1-A(p_0)\)=-{ie^2\over (2\pi)^3}\int dk_0{A(k_0)k_0\over A^2k^2_0-B^2}
D_{00}^0(k_0-p_0),
\label{eqA}
\next
B(p_0)=-{ie^2\over (2\pi)^3}\int dk_0{B(k_0)\over A^2k^2_0-B^2}D_{00}^0(k_0-p_0),
\label{eqB}
\ea
Where $A,B$ are functions of one variable $p_0$.
Transforming this into the Euclidean region ($p_0\to i~p$), we obtain:
\ba
B(p)={e^2\over (2\pi)^3}\int dk{B(k)\over A^2k^2+B^2}D(k-p),
\label{euclB}
\next
p(1-A)={e^2\over (2\pi)^3}\int dk{Ak\over A^2k^2+B^2}D(k-p),
\label{euclA}
\next
D(k)=-\int d^2p_\perp e^{-l^2p^2_\perp/2}D_{00}(ik,p_\perp).
\label{euclD}
\ea

\chapt{2.The vacuum polarization in \QED in an external magnetic field.}

Let us investigate the influence of the strong magnetic field on the vacuum 
polarization. For this purpose let us write the SD equation for 
the polarization tensor: 
\ba
\Pi^{\mu\nu}(x,x')=-ie^2~Tr\int d^3y~d^3z~G(z,x)\gamma^\mu
G(x,y)\Gamma^\nu(y,z|x'),
\label{PhotonSD}
\ea
where  $\Gamma^\nu$ is the full vertex, $G$ is the full electron 
propagator. Since in the strong magnetic field the full 
propagator has the structure (\ref{MatrG}) and
\ba
{1-i\g^1\g^2\over 2}\g^\mu{1-i\g^1\g^2\over 2}=
\g^0 \delta^{\mu,0}{1-i\g^1\g^2\over 2}~~,~~
\Pi^{\mu\nu}=\Pi^{\nu\mu},
\label{Vanish}
\ea
then only $\Pi^{00}$ does not vanish.
Because of $k_\mu\Pi^{\mu\nu}=k_0\Pi^{00}=0$, which is a consequence 
of the gauge invariance, $\Pi^{\mu\nu}(k)=0$. 
Thus the LLL does not contribute to the polarization tensor.
It means that photons become almost free in very strong magnetic field.
The contribution of the next levels into the $\Pi^{00}$ is
 $\sim 1/\sqrt{e|B|}$  (see appendix).

The suppression of the vacuum polarization is connected with 
absence of the longitudinal (with respect to the magnetic field)
components of the polarization tensor in \QED.
In general it takes the form (restrictions are put on by the 
gauge invariance and by the presence of only one singled out 
direction):
\ba
\Pi_{\mu\nu}=\Pi_0\(g_{\mu\nu}k^2-k_\mu k_\nu\)+
\Pi_\perp\(-g^\perp_{\mu\nu}k_\perp^2-k^\perp_\mu k^\perp_\nu\)+
\Pi_\parallel\(g^\parallel_{\mu\nu}k_\parallel^2-
k^\parallel_\mu k^\parallel_\nu\)
\label{generalPi}\next
k_\perp^2=k_1^2+k_2^2~,~k_\parallel^\mu=\delta^{\mu 0}k^0~~,~~
g_\parallel=\(\matrix{1&0&0\cr 0&0&0\cr 0&0&0}\)~~,~~
g_\perp=\(\matrix{0&0&0\cr 0&-1&0\cr 0&0&-1}\).
\label{gperpar}
\ea
The last term in (\ref{generalPi}), which grows with 
the increase of the field in four--dimensional case \cite{nuclph}, vanishes 
in \QED. Thus, the contribution of the vacuum polarization to the
effect of a mass generation is small when $e^2\ll \sqrt{eB}$.
Further it will be taken into account in the following way:
\ba
D_{00}\approx (1-{k_0^2\over k^2}){1\over k^2 (1+\nu_0 Ne^2\ell)}+\lambda{k_0^2\over k^4}~,~~\nu_0=0.14037
\label{vkad}
\ea

\chapt{3.The dynamical mass of fermions in the strong magnetic field.}

Let us analyse equations (\ref{euclB},\ref{euclA}).
It is convinient to rewrite them in the dimensionless form:
\ba
t(1-A(t))&=&\alpha_0\int ds {sA(s)\over A^2s^2+M^2}U(s-t),
\label{equA}\\
M(t)&=&\alpha_0\int ds{M(s)\over A^2s^2+M^2}U(s-t),
\label{equB}\next
M(s)\equiv\ell B(\ell k)~,~A(s)\equiv A(\ell k)~,~\alpha_0=
{e^2\ell\over 4\pi^2(1+\nu_0 Ne^2\ell)}~,\nonumber\next
U(s)=e^{s^2/2}\int\limits_{|s|}^\infty dt {e^{-t^2/2}\over t}
-e^{s^2/2}s^2\int\limits_{|s|}^\infty dt {1-\lambda (1+\Pi(0))\over t^3}e^{-t^2/2}.
\label{us}
\ea
Further it will be shown that only infrared region is responsible 
for the mass generation, therefore we need to know only
infrared behavior of the  $U$.
The first term in (\ref{us}) has the logarithmic
singularity near $s=0$.
The second term is connected with a gauge and is nonsingular, therefore
its contribution is small.
This fact reflects approximate gauge invariance of the ladder 
approximation. Further only singular term will be considered:
$U(s)\approx -ln(s)$.
Let us suppose that  $B(p^2)$ decreases as function of $p^2$,$A(p^2)$ 
tends to the finite value $A_\infty$ when $p\gg B(0)$.
Let us find $A_\infty$. If we differentiate (\ref{equA}) and
omit derivative of $A$ at $s\gg M(0)$, we obtain:
\ba
1-A(t)&=&\alpha_0\int ds {sA(s)\over A^2s^2+M^2}{d\over dt}U(t-s)
\nonumber\\
&\rightarrow&
\alpha_0 \int ds {sA(s)\over A^2s^2+M^2}{d\over dt}U(t)=0.
\label{Ainfty}
\ea
And so  $A_\infty=1$.
To estimate $A(0)$ let us make the next approximation:
\ba
{1 \over A^2p^2+B^2}\approx{1\over p^2+B^2(0)},
\label{approximation}
\ea
which works well both for $p\gg B(0)$ and 
for $p<B(0)$. Then after differentiating the expression in (\ref{equA}) at $t=0$
 we get:
\ba
1-A(0)\approx \alpha_0\int {ds~A(s)\over s^2+M^2(0)}\approx
A(0){\alpha_0\pi\over M(0)}~~,~~A(0)\approx \(1+{\alpha_0\pi\over M(0)}\)^{-1}.
\label{ANull}
\ea
\def\Shro{Shr$\ddot o$dinger~}
We used infrared asymptotics of $U$ and the fact 
that (\ref{approximation}) behaves as $\delta$-function.
Let us find $M(0)$, it defines a fermion dynamical mass.
At first let us prove that a dynamical mass is generated for arbitrary
$e^2~,~N$, if $\ell m_{dyn}\ll 1$. Approximate equation 
\ba
M(t)=\alpha_0\int ds{M(s)\over s^2+M^2(0)}U(s-t).
\label{apprB}
\ea
can be reduced to the \Shro equation as a result of transformation:
\ba
V(x)=-\alpha_0\int ds e^{isx}U(s)~,~
\Psi(x)={1\over 2\pi}\int ds~e^{isx}{M(s)\over s^2+M^2(0)},
\label{peretv}
\next
{d^2\over dx^2}\Psi+(E-V)\Psi=0~~,~~E=-M^2(0).
\label{shred}
\ea
Using the first (singular) term in (\ref{us}), we obtain 
\ba
V(x)&=&-{\alpha_0 \pi^{3/2}\over \sqrt{2}}e^{x^2/2} Erfc\({x\over \sqrt{2}}\)~,~
Erfc(x)={2\over \sqrt{\pi}}\int\limits_x^\infty e^{-t^2}dt,
\label{Vcoor}\\
V(x)&=&-{\alpha_0\pi\over x}~~,~~x\gg 1.
\nonumber
\ea
Therefore the potential is longrange.
With the help of variational method,
\footnote[1]{We can choose $exp(-\varepsilon x)$ as a trial function.}
it is easy to show that equation (\ref{shred}) has eigenvalues
in discrete spectrum if $U(s)>0$ when $s\to 0$. 
In our case it means that mass is generated for arbitrary $e^2~,~N$ if
$\ell m_{dyn}\ll 1$. Let us find $M(0)$.
If we put $s=0$ in (\ref{apprB}), we obtain
\def\a{\alpha_0}
\ba
M(0)=-\alpha_0\int ds{M(s)\over s^2+M^2(0)}ln|s|.
\label{trEqu}
\ea
Because of $M(0)\ll 1$ and the expression to be integrated is 
singular near s=0 then the infrared region gives the main contribution 
to the integral. Therefore we can replace $M(s)$ 
in the right-hand side to the $M(0)$.
Then $M(0)\approx -\pi\a ln~M(0)$. 
The approximate solution of this equation is
\footnote[2]{by using iterations.} 
\ba
M(0)\approx - \pi\a \(ln(\pi\a)+ln(-ln(\pi\a))\).
\label{prosto}
\ea

Then we obtain the expression for the dynamical mass of fermions:
\ba
m_{dyn}&\approx&\sqrt{e|B|}{M(0)\over A(0)}\nonumber\\
&\approx&-{e^2\over 4\pi(1+\nu_0 Ne^2\ell)}
\(1-{1\over ln~\a\pi}\) 
\(ln(\a\pi)+ln(-ln(\a\pi))\)
\label{mass}\\
\a&=&{e^2\ell\over 4\pi^2(1+\nu_0 Ne^2\ell)}.
\nonumber
\ea
  
It is interesting to consider the case of the weak field
in comparison with charge: $e^2\gg \sqrt{eB}$. In this case 
\ba
m_{dyn}=-{1\over 4\pi\nu_0 N}\sqrt{e|B|}\(1-{1\over ln~\a\pi}\)
\(ln(\a\pi)+ln(-ln(\a\pi)\)
~,~\a={1\over 4\pi^2\nu_0 N}.
\label{SmallB}
\ea

When $N\gg 1$ the condition $\ell m_{dyn}\ll 1$ is true, i.e.
magnetic field can be turned off without breaking
of the approximation $e^2\ell\ll 1$. It is in agreement with the fact that 
without a magnetic field fermions are massless when $N\gg 1$.

This solution for $m_{dyn}$ is non--analytic by the
coupling constant and it can't be evaluated in
the perturbation theory.
It is in the full agreement with statements in \cite{simon},\cite{klaus}
about analytic dependence of the ground state energy
in the one--dimensional \Shro equation with
the long--range potential ($V(x)\sim 1/x~,~x\gg 1$).
Non--analytical dependence of
$m_{dyn}(\a)$ essentially differs \QED and
 2+1 dimensional NJL where dependence of the $m_{dym}$ on the
coupling constant is analytical: 
\ba
m_{dyn}=|eB|{N G\over 2\pi}~~,~~G\to 0,
\label{MNJL}
\ea
where  $N$ is number of flavors, $G$ is a coupling constant in the NJL model. 
The analytical dependence of $m_{dyn}(G)$ is due to the short--range
character of the potential $V\sim \delta(x)$ in this model.
In ref.  \cite{klaus} it is shown that the dependence
of the ground state energy on the coupling constant in the 
one--dimensional \Shro equation 
\ba
{d^2\Psi\over dx^2}+(E-\alpha V)\Psi=0
\nonumber
\ea
is analytical if the next condition takes place:
\ba
\left|\int\limits_{-\infty}^{\infty}(1+|x|)V(x)dx\right|<\infty~~,V(x)\leq 0.
\label{acondition}  
\ea

This condition is true for the NJL model and broken for \QED
 
\chapt{Conclusion.}

  With the help of the SD equation the problem of the 
dynamical mass generation in \QED in presence of the
strong magnetic field was considered. The main
effects that are due to the magnetic field are:
\begin{enumerate}
    \item Fermions become massive for arbitrary number of
          flavors in the system although without an external
          field the mass is generated only for $N<N_c\approx 4.32$
	  This is connected in its turn with the dimensional reduction
	  ($D\to D-2$), which takes place in this problem.

    \item In the case of a very strong magnetic field
	  the vacuum polarization almost does
          not contribute to the result.
          Photons become almost free in strong magnetic
          field. It is connected with absence of the 
	  longitudinal (with respect to the magnetic field) directions.
	  This effect essentially differs \QED from \QEDR.

    \item Strong magnetic field shifts the mass generation to
	  the infrared region ($p\leq m_{dyn}$). In the \QED without external magnetic field
	  the middle region $m_{dyn}<p<N~e^2/8$ is responsible
	  for the mass generation.

    \item The dependence of the $m_{dyn}$ on the coupling constant
	  is non--analytical in contrast to the NJL model.

\end{enumerate}

\chapt{Acknowledgments.}

Author is greatful to V.P. Gusynin for the valuable advices, to
V.A. Miransky for the fruitful discussions about the role 
of the vacuum polarization, to I.A. Shovkovy for the
checking and discussions of some results.
This work was in part supported by the International Soros
Science Education Program (ISSEP), grant GSU052402.

\pagebreak
\chapt{Appendix.}
\chapl{The polarization tensor in presence of a constant magnetic field.}

  The polarization tensor can be evaluated in the second order 
of the perturbation theory using the method of \cite{Berlin}.
\ba
\Pi_{\mu\nu}(k)=-ie^2tr\<\g_\mu S(p)\g_\nu S(p-k)\>~~,~~\<...\>=\int {d^3 p\over (2\pi)^3}...
\label{Pidef}
\ea

$S$ is the free fermion propagator. Let us represent
it in the next form:
\ba
S(k)&=&-i\intz ds~exp\({-is\(m^2-k_0^2+k^2_\perp {tg~z\over z}\)}\)
\nonumber\\
&\cdot&\(k^\mu\g_\mu+m+(k^2\g^1-k^1\g^2)tg~z\)\(1+\g^1\g^2 tg~z\)=
\nonumber\next
=-i\intz ds~exp\({-is\(m^2-k_0^2+k_\perp^2{tg~z\over z}\)}\)
\( (k^0\g^0+m){e^{i\sigma_3z}\over cos~z}+{\g k_\perp\over cos^2z}\)
~,
\label{bigFree}\\
z=eBs.
\nonumber
\ea

After some computations we have obtained the expression 
for the polarization tensor:
\def\inr{e^{i\pi/4}{e^2\over 4\pi\sqrt{\pi}}\intz
{ds\over \sqrt{s}}\int\limits_{-1}^1{dv\over 2}e^{-is\varphi_0}}
\def\inre{{e^2\over 4\pi\sqrt{\pi}}\intz
{ds\over \sqrt{s}}\int\limits_{-1}^1{dv\over 2}e^{-s\varphi}}
\ba
\Pi_{\mu\nu}(k)=\Pi(k)\(\gmn k^2-k_\mu k_\nu\)+
\Pi_\perp(k)\(-\gper k_\perp^2-k^\perp_\mu k^\perp_\nu\),
\label{gradinvPi}\next
\Pi&=&\inr\({z~cos~zv-zv~ctg~z~sin~zv\over sin~z}\),
\label{PiTotal}\\
\Pi_\perp&=&\inr {cos~zv-cos~z\over sin^3z}2z-\Pi.
\label{PiPerp}
\ea
The sign $'-'$ before the $g_\perp$ is due to the $k_\perp^2=k_1^2+k_2^2$.
The full photon propagator is equal to:
\ba
D_{\mu\nu}&=&{1\over k^2(1+\Pi)}\(\gpar-{k^\perp_\mu k^\perp_\nu\over k^2_\perp}
-{k_\mu k_\nu\over k^2}\)\nonumber\\
&+&{1\over k^2+k^2\Pi-k_\perp^2\Pi_\perp}\(g^\perp_{\mu\nu}+
{k^\perp_\mu k^\perp_\nu\over k^2_\perp}\)+\lambda{k_\mu k_\nu\over k^4}.
\label{FullDmunu}
\ea 
 In Euclidean notation the quantities  $\Pi,\Pi_\perp$ take the form:
\ba
\Pi&=&\inre\({z~ch~zv\over sh~z}-{zv~cth~z~sh~zv\over sh~z}\),
\label{EuclPiTotal}\\
\Pi_\perp&=&\inre {ch~z-ch~zv\over sh^3z}2z-\Pi,
\label{EuclPiPerp}\\
\varphi&=&m^2+{1-v^2\over 4}k_3^2+{ch~z-ch~zv\over 2z~sh~z}k^2_\perp
~~,~~k_3=ik_0.
\label{EuclPhi}
\ea
Both integrals are finite for $|B|>0$. Because we need only 
(\ref{EuclPiTotal}), let us investigate it in detail. 
For zero momentum
\ba
\Pi(m)&=&{e^2\ell\over 4\pi\sqrt{\pi}}\intz {dz\over \sqrt{z}}e^{-\ell^2 m^2 z}
\({cth~z\over z}-{1\over sh^2 z}\)
\label{PiM}\\
\Pi(0)&\approx&0.14037 e^2\ell=\nu_0~e^2\ell
\nonumber
\ea
While $p_3^2~$ increases,$~p_\perp^2$ (\ref{EuclPiTotal}) 
slowly decreases. Because the leading contribution is
given from the infrared region, we can replace $\Pi(p)$ to the $\Pi(0)$.
If there are $N$ flavors in the system then all $\Pi$ must be
multiplied on $N$ because  $N$ equalent diagramms 
contribute to (\ref{Pidef}).

\pagebreak
\bib
\ce{\bf Literature.}
\baselineskip=21pt

\bibitem{nuclph} {\it V.P.Gusynin, V.A.Miransky,I.A.Shovkovy}//
Nucl.Phys.1996,v.B462,p.249

\bibitem{PhLet}{\it V.P.Gusynin, V.P.Miransky, I.A.Shovkovy}//
Phys.Rev.Lett.1994,v.73,p.3499

\bibitem{mqed} {\it V.P.Gusynin, V.A.Miransky, I.A.Shovkovy}
//Phys.Rev.1995,v.D52,p.4747

\bibitem{mnjl} {\it V.P.Gusynin, V.A.Miransky, I.A.Shovkovy}//
Phys.Rev.1995,v.D52,p.4718

\bibitem{applq}
{\it T.W.Appelquist, M.Bowick, D.Karabali, L.C.R.Wijewardhana}\\ //Phys.Rev.1986,v.D33,p.3704\\  
{\it T.W.Appelquist, D.Nash, L.C.R.Wijewardhana}
//Phys.Rev.Lett.1988,v.60,p.2575\\
{\it D. Nash}// Phys.Rev.Lett.1989,v.62,p.3024

\bibitem{QED3} {\it V.P.Gusynin, A.H.Hams, M.Reenders}//Phys.Rev.
1996,v.D53,p.2227

\bibitem{physMean} {\it N. Dorey, E. Mavromatos}//Nucl.Ph.1992,v.B368,p.614\\
{\it I.V. Krive, A.S. Rozhavsky}//Sov.Phys.Usp,1987,v.30,p.370

\bibitem{Nuovo} {\it P.I. Fomin, V.P. Gusynin, V.A. Miransky, Yu.A. Sitenko}\\
//Riv.Nuovo Cimento,1983,v.6,p.1

\bibitem{Berlin} {\it W.Dittrich and M.Reuter}//{\bf Effective lagrangians
in Quantum\\ Electrodynamics}, Springer-Verlag,Berlin,1985

\bibitem{decomp} {\it A.Chodos, A.Everding, D.A.Owen}//Phys.Rev.
1990,v.D42,p.2881

\bibitem{shwing} {\it J.Schwinger}// Phys.Rev.1951,v.82,p.664

\bibitem{simon} {\it B.Simon}//Ann.Phys.1976,v.97,p.279

\bibitem{klaus} {\it M.Klaus}//Ann.Phys.1977,v.108,p.288-300

\finbib

\pagebreak


\end{document}